\documentstyle[preprint,aps]{revtex}
\begin{document}
\draft

\title{\Large\bf The transverse polarized structure function in DIS \\
	and chiral symmetry breaking in QCD}

\author{{\bf Wei-Min Zhang and A. Harindranath\thanks{On leave of
absence from Saha Institute of Nuclear Physics, Sector 1, Block AF,
Bidhan Nagar, Calcutta 700064 India.}} \\
	International Institute of Theoretical and Applied Physics\\
	123 Office and laboratory, Iowa State University, Ames, Iowa 50011}
\date{June 9, 1996}

\maketitle

\begin{abstract}
We study the polarized structure functions in QCD. 
We show that $g_T$ which probes helicity flip interactions  
in hadrons on the light-front indeed measures the QCD 
dynamics of chiral symmetry breaking. The relation between 
chiral symmetry breaking and the observed $g_2$ data is
explored.
\end{abstract}

\vspace{0.5in}


%\vspace{0.5in}

Polarized structure functions, in particular, the transverse
polarized structure function $g_T=g_1+g_2$, have recently 
received much theoretical and experimental attention. Preliminary 
extraction of $g_2$ has been made in the deeply inelastic 
scatterings (DIS) by the SMC experiment in CERN and
the E143 experiment in SLAC very recently \cite{E143}. 
Unlike the longitudinal polarized structure function 
$g_1$ which measures the quark helicity distribution 
in the longitudinal polarized hadrons, the physical 
interpretation of $g_2$ is not simple\cite{Review}. Much 
theoretical work on $g_2$ is currently concentrated on 
the questions whether $g_2$ can still be described 
approximately by parton distributions\cite{Ralston},
and whether it is a relatively good approximation to 
predict $g_2$ from $g_1$ via the Wandzura-Wilczek 
relation \cite{WW,Roberts} or whether the quark-gluon 
coupling can provide a significant contribution to 
$g_2$ \cite{Jaffe}. 

In this letter we show that $g_T$ probes the light-front
helicity flip interactions in hadrons. The helicity
flip on the light-front is the manifestation of chiral 
symmetry breaking in QCD. Therefore, 
$g_T$ constitutes a direct measurement of QCD chiral 
symmetry breaking. We also explore the explicit relation 
between dynamical chiral symmetry breaking and the observed 
$g_2$ data.

The polarized structure functions in DIS are defined from the
antisymmetric part of hadronic tensor,
\begin{equation}
	W^{\mu \nu}_A = -i \epsilon^{\mu \nu \lambda \sigma}q_\lambda
		\Bigg\{ {S_\sigma \over \nu }(g_1(x,Q^2) +g_2(x,Q^2)) -
		P_\sigma {S\cdot q\over \nu^2} g_2(x,Q^2) \Bigg\}
\end{equation}
where $P$ and $S$ are the target four-momentum and polarization 
vector respectively ($P^2=M^2, S^2=-M^2, S\cdot P=0$), and $q$ is 
the virtual-photon four momentum ($Q^2=-q^2$,  $\nu = P\cdot q$,
$x={Q^2\over 2\nu}$). 
On the other hand, the hadronic tensor is related to 
the forward virtual-photon hadron Compton scattering amplitude:
\begin{equation}	\label{vpfc}
	W^{\mu \nu} = {1\over 4\pi}{\rm Im} T^{\mu \nu}~~,~~~~ 
		T^{\mu \nu} = i \int d^4\xi e^{iq\cdot \xi}
		\langle PS | T(J^\mu(\xi) J^\nu(0) |PS \rangle .
\end{equation}

We first derive the hadronic matrix element expression for $g_1$ 
and $g_2$.  We shall not begin with the assumptions that have 
been used in the previous derivations, such as the zero quark 
mass and zero transverse quark momentum limits in the naive quark
model, the free quark field assumption in the impluse approximation, 
and even the factorization assumption in the collinear expansion 
approximation.

We begin with the ${1\over q^-}$ expansion of $T^{\mu \nu}$ \cite{Jackiw}, 
\begin{equation} \label{lfcc}
	q^- T^{\mu \nu} = \int d\xi^- d^2 \xi_\bot e^{iq\cdot \xi}
	  \langle PS | [J^\mu(\xi), J^\nu(0)]_{\xi^+=0}| PS\rangle
		+ O\Big({1\over q^-}\Big),
\end{equation}
where $q^-=q^0 - q^z$. For large $Q^2$ and $\nu$ limits in 
DIS which correspond to large $q^-$, we ignore the 
contributions from terms of the order ${1\over q^-}$ in eq.(\ref{lfcc}).  
What remains is proportional to a light-front current commutator 
which can be computed directly from QCD (where QCD is 
quantized on the light-front time surface $\xi^+=\xi^0+\xi^3=0$ 
with the light-front gauge $A_a^+=0$ \cite{KS,Brodsky,Zhang93}). 
Then we can show 
\begin{eqnarray}
	g_1(x,Q^2) &=& {1\over 4\pi S^+} \int_{-\infty}^{\infty}
		d\eta e^{-i\eta x} \langle P S| \psi_+^\dagger (\xi^-)
		{\cal Q}^2\gamma_5 \psi_+(0) + h.c. |PS \rangle , \label{g1}\\
	g_T(x,Q^2) &=& {1\over 8\pi (S_\bot-{P_\bot\over P^+}S^+)} 
		\int_{-\infty}^{\infty} d\eta e^{-i\eta x} \langle PS | 
		\Big(O_m + O_{k_\bot} + O_g \Big)  + h.c |PS \rangle  
		\nonumber \\ &=& g_T^m(x,Q^2) + g_T^{k_\bot}(x,Q^2) + g_T^g
			(x,Q^2), \label{g2} 
\end{eqnarray}
where the parameter $\eta \equiv {1\over 2}P^+ \xi^-$ with $\xi^-$ 
being the light-front longitudinal coordinate, and ${\cal Q}$  
the quark charge operator. We have also defined $\psi_+ \equiv {1\over 2}
\gamma^0\gamma^+ \psi$ which is the light-front quark field, and
$g_T \equiv g_1 +g_2$. The operators in eq.(\ref{g2}) are given 
as follows:
\begin{eqnarray}
	&& O_m=m \psi_+^\dagger (\xi^-) {\cal Q}^2 \gamma_\bot 
		\Bigg({1 \over i \roarrow{\partial}^+} - {1\over 
		i \loarrow{\partial}^+} \Bigg)\gamma_5 \psi_+(0),
			\nonumber \\
	&& O_{k_\bot}= -\psi_+^\dagger (\xi^-){\cal Q}^2\Bigg(\gamma_\bot
		{1\over\roarrow{\partial}^+}{\not \! \roarrow{\partial_\bot}} 
		+ {\not \! \loarrow{\partial_\bot}}{1\over \loarrow{
		\partial}^+}\gamma_\bot + 2{P_\bot \over P^+}\Bigg) 
		\gamma_5\psi_+(0), \nonumber \\
	&& O_g = g\psi_+^\dagger(\xi^-){\cal Q}^2 \Bigg({\not \! \!
		A_\bot}(\xi^-){1\over i\loarrow{\partial}^+}\gamma_\bot 
		- \gamma_\bot {1\over i\roarrow{\partial}^+}{\not \! \! 
		A_\bot}(0) \Bigg)\gamma_5 \psi_+(0) ~ \label{go}
\end{eqnarray} 
and $m$ and $g$ are the quark mass and quark-gluon coupling 
constant in QCD, and $A_\bot=A^a_{\bot}T_a$ the transverse 
gauge field.

Since we work in the light-front gauge, the operators in 
eqs.(\ref{go}) are well-defined. Eqs.(\ref{g1}-\ref{g2}) 
are also the general expressions for the 
target being in any arbitrary frame $\{P^\mu \}$.  By using 
the light-front decomposition, $\psi = \psi_+ + \psi_-$, 
$\psi_- = {\gamma^0 \over i \partial^+}({\not \! \! D_\bot} 
+m)\psi_+$, we can formally rewrite eqs.(\ref{g1}-\ref{g2}) 
in familiar expressions, 
\begin{eqnarray}
	&& g_1(x,Q^2) = {1\over 8\pi S^+} \int_{-\infty}^{\infty}
		d\eta e^{-i\eta x} \langle PS |~ \overline{\psi}(\xi^-)
		{\cal Q}^2 \gamma^+ \gamma_5 \psi(0)+ h.c. |PS \rangle, \\
	&& g_T(x,Q^2) = {1\over 8\pi (S_\bot-{P_\bot\over P^+}S^+ )} 
		\int_{-\infty}^{\infty} d\eta e^{-i\eta x} \langle PS |
		~ \overline{\psi}(\xi^-){\cal Q}^2 \Bigg(\gamma_\bot 
		-{P_\bot\over P^+}\gamma^+ \Bigg) \gamma_5 \psi(0) + h. c. 
		|PS \rangle .
\end{eqnarray}
However, the physical picture is clearer in the expressions
eqs.(\ref{g1}-\ref{g2}), where as we can see $g_T$ contains 
explicitly the contributions associated with the quark mass, 
quark transverse momentum and quark-gluon coupling. Note that 
$g_2$ cannot be directly computed in the physical basis.  
We can extract $g_2$ from 
$g_1$ and $g_T$, only the latter two structure functions can be 
directly calculated and experimentally measured in the longitudinal 
and transverse polarized targets, $|P\lambda\rangle$ and 
$|PS_\bot \rangle$, respectively.
 
Also note that eqs.(\ref{g1}-\ref{g2}) are expressed in terms 
of equal {\it light-front} time matrix elements. It is most 
convenient to analyze these matrix elements in light-front Fock 
space expansion. The results are
\begin{eqnarray}
	g_1(x,Q^2) &=& {1\over 2} \sum_i e_i^2 ~\Delta q^L_{i}
		(x,Q^2) \label{g1d} \\
	g_T(x,Q^2) &=& {1\over 2xM} \sum_i e_i^2 \Bigg\{ m_i  
		\Delta q^T_{i}(x,Q^2) + \Delta {\cal K}_i(x,Q^2)
		+ g {\cal T}^g_i(x,Q^2) \Bigg\}, \label{gtd}
\end{eqnarray}
where $i$ is the flavor index, the notation $\Delta A_i \equiv A^+_{i}
-A^-_{i} + \overline{A}^+_i-\overline{A}^+_i $, 
\begin{eqnarray}
	q_{i}^{L\pm} (x,Q^2) &=&  \int {d^2 k_\bot\over 2(2\pi)^3}
		\langle P \lambda |b^\dagger_i(x,k_\bot,\pm \lambda)
		b_i(x,k_\bot,\pm \lambda) |P\lambda \rangle , 
		\label{qhd} \\
	q^{T\pm}_{i} (x,Q^2) &=& \int {d^2 k_\bot\over 2(2\pi)^3}
		\langle P S^1 |b^\dagger_i(x,k_\bot,\pm s^1)
		b_i(x,k_\bot,\pm s^1) |P S^1 \rangle , \\
	{\cal K}^{\pm}_i (x,Q^2)&=& \int {d^2 k_\bot\over 2(2\pi)^3}
		\kappa^1 \langle P S^1 |b^\dagger_i(x,k_\bot,
		\pm \lambda) b_i(x,k_\bot,\pm \lambda) |P S^1 
		\rangle , \label{tmd}
\end{eqnarray}
and  $\overline{q}_i^\pm$ and $\overline{\cal K}_i^\pm$ have similar
form for antiquarks. In eqs.(\ref{qhd}-\ref{tmd}), $\lambda$ is the 
light-front helicity (the eigenvalue of the Pauli matrix $\sigma_z$). 
Without loss of generality, we have also taken the transverse 
polarization of the target in the $x$-direction: $S^1$, and 
$\kappa^1=k^1_\bot - xP^1_\bot$ is the $x$-component of the 
relative transverse quark momentum, while ${\cal T}^g_i$ has no 
simple expression.

With the light-front quantization being utilized\cite{Hari}, 
the physical interpretation of the above results becomes rather 
simple. The $g_1$ is purely determined by the quark and antiquark
helicity distribution $\Delta q^L_{i}$. 
The transverse polarized structure function $g_T$ contains 
three contributions, as we have mentioned. The contribution 
associated with quark mass $g_T^m$ is proportional to the 
transverse polarized distribution $\Delta q^T_{i}$. Apparently, 
the contribution associated 
with transverse quark momentum $g_T^{k_\bot}$ is proportional 
to $\Delta {\cal K}_i$ which measures averages of the
transverse momentum $\kappa_\bot$ of quarks and antiquarks 
with helicity up and down in the 
transverse polarized target. Besides, $g_T$ also includes 
the contribution $g_T^g$ from the quark-gluon coupling, which is
proportional to ${\cal T}_g$ and describes 
dynamical processes of a parton emitting and absorbing a gluon. 
At this step, formally the later two contributions in $g_T$ 
do not have a simple parton picture, and they are the most 
interesting quantities in the current study of $g_2$.  It has 
been suggested that the contribution proportional to quark mass 
is small since the current quark mass is  small. 
Therefore, the later two contributions, $g_T^{k_\bot}$ and 
$g_T^g$, appear to be dominant in the transverse polarized 
structure function. 

However, we find that, first of all, the main contributions from 
$g_T^{k_\bot}$ and $g_T^g$ have indeed the simple parton picture 
just as $g_T^m$ but they do not manifest at the tree level of QCD. 
Secondly, the nontrivial dynamics determined by $g_T$ comes 
from the dynamical chiral symmetry breaking of nonperturbative QCD.
To clearly see what is the physical origin of such $g_T^{k_\bot}$ 
and $g_T^g$ contributions and how dynamical chiral symmetry breaking
dominates the physics of $g_T$, we must have further knowledge on 
the target bound state. The target state with transverse polarization 
in the $x$-direction can be expressed as a combination of the 
helicity up and down states: $|P S^x\rangle ={1\over \sqrt{2}}
(|P\uparrow\rangle \pm |P\downarrow\rangle)$ for $S^x=\pm M$. 
Then we have
\begin{equation}
	g_T(x,Q^2) = {1\over 8\pi M} \int_{-\infty}^{\infty} 
		d\eta e^{-i\eta x} {1\over 2} \sum_\lambda
		\langle P \lambda |\Big(O_m + O_{k_\bot} + O_g \Big)  
		+ h.c | P\!-\!\lambda \rangle .  \label{helib} 
\end{equation}
This shows that $g_T$ measures the helicity flip dynamics of
hadrons. 

So far, we have not specified the general structure of $|PS \rangle$. 
Generally, on the light-front,
\begin{equation}
	| PS \rangle = \sum_{n,\lambda_i}\int' {dx_i d^2\kappa_{\bot i} 
		\over 2 (2\pi)^3 }| n, x_iP^+,x_iP_{\bot}+ \kappa_{\bot i},
		\lambda_i \rangle \Psi^S_n (x_i,\kappa_{\bot i},\lambda_i), 
		\label{lfwf}
\end{equation}
where $|n, x_i P^+, x_i P_{\bot} + k_{\bot i}, \lambda_i \rangle$ 
is a Fock state with $n$ constituents, $\int'$ denotes the integral 
over the space $(x_i, \kappa_{\bot i})$ with $\sum_i x_i = 1$ and 
$\sum_i \kappa_{\bot i} = 0$, where $x_i = { k_i^+ \over P^+}$, 
$\kappa_{\bot i} = k_{\bot i} - x_i P_{\bot}$, and $k_i^+, k_{\bot i}$ 
are the longitudinal and transverse momentum of the $i$-th 
constituent with helicity $\lambda_i$. The amplitude $\Psi^S_n(x_i,
\kappa_{\bot i},\lambda_i)$ is determined by the QCD eigenvalue 
equation $H^{LF}_{QCD} |PS \rangle = { P_{\bot}^2 + M^2 \over P^+ } 
|PS \rangle$ which can be explicitly written as \cite{Brodsky}
\begin{equation}
	\Big(M^2 - \sum_i { \kappa_{i\bot}^2 + m_i^2 \over x_i} \Big)
		\left(\begin{array}{c} \Psi_{qqq} \\ \Psi_{qqqg}
		\\ \vdots \end{array} \right) 
		  = \left( \begin{array}{ccc} \langle qqq
		| H_{I} | qqq \rangle & \langle qqq | H_{I} 
		| qqq g \rangle & \cdots \\ \langle qqq g
		| H_{I} | qqq \rangle & \cdots & ~~  \\ \vdots &
		&  ~~ \end{array} \right) \left(\begin{array}{c} 
		\Psi_{qqq} \\ \Psi_{qqqg} \\ \vdots \end{array} 
		\right), \label{lfbe}
\end{equation}
where $H^{LF}_{QCD} = H_0+H_{I}$. Note that $\Psi^S_n(x_i,\kappa_{
\bot i},\lambda_i)$ is only a function of $(x_i, \kappa_{\bot i})$ 
as a result of the kinematic boost symmetry in light-front theory.

A complete understanding of $g_T$ depends of course on the 
solution of eq.(\ref{lfbe}). For some of the approaches to 
solve the above bound state equation see refs.\cite{dlcq,tm,Wilson94}. 
But here without explicitly solving the nucleon bound state
from eq.(\ref{lfbe}), we show that the 
dominant contributions from $g_T^{k_\bot}$ and $g_T^g$ are 
proportional to quark mass and the transverse polarized 
distribution $\Delta q^T_{i}(x,Q^2)$. 

From eq.(\ref{lfbe}), as we see the higher Fock states in 
the hadronic bound states are generated by the interaction
part of QCD Hamiltonian.
For large $Q^2$, we can rewrite the state eq.(\ref{lfwf}) 
as the bound state $|\Phi(P,S,\mu)\rangle$ at hadronic scale 
$\mu \sim M $  plus the radiative corrections
from the high energy $H_{I}$ in the $\xi^+$-ordering perturbative 
expansion:
\begin{equation}
	|PS\rangle = \sum_{n=0}^\infty \Bigg({H_{I}\over P^- 
		-H_0}\Bigg)^n |\Phi(P,S,\mu)\rangle ,
\end{equation}
where all quarks and gluons in $H_{I}$ are restricted 
to $ \mu^2 \leq \kappa^2_\bot \leq Q^2$.  Then,
\begin{eqnarray}
	g_T(x,Q^2) &=& {1\over 8\pi (S_\bot-{P_\bot\over P^+}
		S^+) } \int_{-\infty}^{\infty} 
		d\eta e^{-i\eta x} \sum_{n_1,n_2} \langle P,S,\mu|
		n_1\rangle \langle n_1|\sum_{n=0}^\infty \Bigg(
		{H_{I}\over P^- -H_0}\Bigg)^n  \nonumber \\
		&& ~~~~~~~~ \times \Bigg\{\Big(O_m + O_{k_\bot} 
		+ O_g \Big)  + h.c \Bigg\}\sum_{n'=0}^\infty 
		\Bigg({H_{I}\over P^- -H_0}\Bigg)^{n'}|n_2\rangle 
		\langle n_2| P,S,\mu \rangle , \label{gfock}  
\end{eqnarray}
where $|n\rangle$ is a simple notation of $|n, x_i P^+, x_i 
P_{\bot} + k_{\bot i}, \lambda_i \rangle$.

We first consider those terms in eq.(\ref{gfock}) with $|n_1\rangle 
=|n_2 \rangle$. This will immediately lead to $g_T \sim \Delta 
q^T_{i}(x,Q^2)$ for large $Q^2$, and the coefficient is determined 
by the matrix element 
\begin{eqnarray}
	&& \langle n_1|\sum_{n=0}^\infty \Bigg({H_{I} \over P^- 
		-H_0}\Bigg)^n \Big(O_m + O_{k_\bot} + O_g \Big) 
		\sum_{n'=0}^\infty \Bigg({H_{I}\over P^- 
		-H_0}\Bigg)^{n'}|n_1\rangle \nonumber \\ 
&& ~~~~~~~~~~~~~~~~~~~\stackrel{{\rm large}~Q^2}{\longrightarrow} 
		\langle 1|\sum_{n=0}^\infty \Bigg({H_{I} 
		\over P^- -H_0}\Bigg)^n \Big(O_m + O_{k_\bot} + O_g 
		\Big) \sum_{n'=0}^\infty \Bigg({H_{I}\over P^- 
		-H_0}\Bigg)^{n'}| 1 \rangle  , \label{sqm}
\end{eqnarray}
here we denoted $|1 \rangle = | y, k_\bot, s_\bot \rangle$ which 
means that we have suppressed the states of all the spectators, 
while $y=k^+/P^+$. 

Without the QCD correction, it is easy to show that only 
the quark mass term contributes to $g_T$ in eq.(\ref{sqm}), 
\begin{equation}
	M_T^m(x,y) =e_q^2 m_q \delta(y-x), 
		~~M_T^{k_\bot}(x,y)=0=M_T^g(x,y),
\end{equation}
where $M_T^i \equiv {1\over 4\pi}\int_{-\infty}^{\infty} d\eta 
e^{-i\eta x} \langle 1 | O_i| 1\rangle$. The physical picture
of this result is as follows. In terms of the helicity basis 
eq.(\ref{helib}), $g_T$ measures helicity flip of quarks. 
The quark mass term $O_m$ already flips the helicity of one quark 
so that its matrix element in eq.(\ref{sqm}) does not vanish. 
But the operator $O_{k_\bot}$ and $O_g$ do not change quark 
helicity of the states, the corresponding matrix elements 
must vanish.

Next, we consider the QCD corrections up to order $\alpha_s$. 
We find that all the three matrix elements in eq.(\ref{sqm}) 
have the nonzero contribution to $g_T$, 
\begin{eqnarray}
	M_T^m(y,x,Q^2) &=&e^2_q m^R_q \Bigg\{\delta(y-x) + {\alpha_s 
		\over 2\pi} C_f \ln{Q^2\over \mu^2} \Bigg[{2 \over 
		y-x} \nonumber \\
	&& ~~~~~~~~~~~~~~~~  -~\delta(y-x) \Bigg({3\over 2} + \int_0^1 dx'
		{1+x'^2\over 1-x'} \Bigg) \Bigg]\Bigg\}, \label{ems}\\
	M_T^{k_\bot}(y,x,Q^2) &=& -e^2_q m^R_q {\alpha_s \over 2\pi} 
		C_f \ln{Q^2\over \mu^2}{(y-x)\over y^2} , \label{gtk} \\
	M_T^g(y,x,Q^2) &=& e_q^2 m^R_q{\alpha_s \over 2\pi} C_f 
		\ln{Q^2\over \mu^2}{\delta(y-x) \over 2}, \label{gtg}
\end{eqnarray}
where $\mu^2 > (m^R_q)^2 $, and $m^R_q$ is the renormalized mass at 
the hadronic scale $\mu$. [The term $\sim {3\over 2} \delta(y-x)$ 
in eq.(\ref{ems}) is a result of replacing the bare quark mass by 
the renormalized one. Note that missing this mass renormalization 
effect will lead to the violation of Burkhardt-Cottingham sum 
rule]. 
%Thus,
%\begin{eqnarray}
%	M_T(y,x,Q^2)&=&M^m_T(y,x,Q^2)+ M^{k_\bot}_T(y,x,Q^2)+ M^g_T
%		(y,x,Q^2) \nonumber \\
%	  &=& e^2_q m^R_q \Bigg\{\delta(y-x) + {\alpha_s \over 2\pi} 
%		C_f \ln{Q^2\over \mu^2} \Bigg[{y^2+2yx-x^2 \over y^2 
%		(y-x)_+} + {1\over 2} \delta(y-x) \Bigg] \Bigg\} . 
%		\label{gt1l}
%\end{eqnarray}
It shows that up to order $\alpha_s$, the matrix 
elements from $O_{k_\bot}$ and $O_g$ in eq.(\ref{sqm}) 
are also proportional to the quark mass and they do provide 
a similar contribution to $g_T(x,Q^2)$ as that of $O_m$. 

What is the physical reason that makes the matrix elements of 
$O_{k_\bot}$ and $O_g$ have nonzero
contribution to $\Delta q^T_i(x)$ when the QCD correction is
considered? The answer comes from the underlying QCD dynamics 
on the light-front. When QCD is quantized on the light-front,
one can find that there is a quark-gluon interaction term in 
the QCD Hamiltonian which is proportional to quark mass 
(see ref.\cite{Zhang93}), 
\begin{equation}
	 -gm_q \psi_+^\dagger\Bigg({\not \! \! A}_\bot
		{1\over i\partial^+} + {1\over i\partial^+}{\not \! 
		\! A}_\bot \Bigg)\psi_+ .  \label{chiral}
\end{equation}
At the canonical level, only this term can flip quark helicities 
in QCD. The nonzero contributions of $g_T^{k_\bot}$ and $g_T^g$ 
arise because the matrix element of
eq.(\ref{sqm}) contains the helicity flip from this mass term 
in $H_{I}$. Therefore, it is this helicity flip interaction 
of QCD that generates the contributions from $g_T^{k_\bot}$ and 
$g_T^g$ that is proportional to $m_q^R$ and $\Delta q^T_i(x,Q^2)$. 

Meanwhile, the matrix elements of $O_{k_\bot}$ and $O_g$ in 
eq.(\ref{gfock}) also have the contributions to $g_T$ that 
are not proportional to the transverse polarized distribution. 
These correspond to the cases where i) although $n_1=n_2$ 
the single quark states of the matrix element in
eq.(\ref{sqm}) are transversely polarized in the opposite 
direction, and ii) $n_1 \neq n_2$ (different by a gluon). The 
corresponding contributions to $g_T$ are proportional to 
the non-diagonal matrix elements given by $\Delta {\cal K}_i$ 
and ${\cal T}_i$ in eq.(\ref{gtd}), respectively. In other 
words, $\Delta {\cal K}_i$ and ${\cal T}_i$  only 
contain the part of the contributions from $g_T^{k_\bot}$ 
and $g_T^g$ that does not have the simple parton picture.

Now, as we see the first term in eq.(\ref{gtd}) that is proportional 
to quark mass contains all the contributions from the three terms
in eq.(\ref{g2}) after we replace the bare quark mass by 
the renormalized one, where the contributions from $g_T^{k_\bot}$
and $g_T^g$ originate from the helicity flip 
quark-gluon interaction in QCD. It is well known that on the 
light-front the helicity is just the chirality. Helicity flip 
corresponds to chiral symmetry breaking on the light-front. 
Thus, only the helicity flip interactions, such as the one 
given by eq.(\ref{chiral}), are responsible for the chiral 
symmetry breaking in light-front QCD.  As it has been pointed out 
\cite{Wilson94}, the light-front QCD vacuum can be simplified in a 
cutoff theory so that the dynamics of the spontaneous chiral 
symmetry breaking in nonperturbative QCD can become an explicit
chiral symmetry breaking by the manifestation of effective
quark-gluon interactions in the QCD Hamiltonian. Any such
interaction that is responsible for the spontaneous chiral
symmetry breaking must be a helicity flip interaction. These 
interactions can contribute to $g_T$ just in the same way as 
the canonical interaction of eq.(\ref{chiral}).  Thus, 
there is a contribution to $g_T$ that arises from the spontaneous
chiral symmetry breaking in nonperturbative QCD. This contribution
can be simply taken into account by requiring that the renormalized 
quark mass parameter does not vanish in the chiral limit. Therefore, 
we can effectively write $m^R_q=m_q^c + \chi_q$, where $m_q^c$
is a current quark mass and $\chi_q$ is associated with 
the spontaneous chiral symmetry breaking in QCD. 

Meanwhile, the transverse polarized distribution $\Delta g^T_i (x)$ 
which has the probabilistic interpretation is proportional to 
the modulus squared of the amplitudes of all the Fock states 
in eq.(\ref{lfwf}). But $\Delta {\cal K}_i$ and ${\cal T}_i$ are the 
off-diagonal matrix elements that are proportional to the 
amplitude mixings with different Fock states. These are smaller 
in comparison to the modulus squared of amplitudes and also 
have potential cancellations between different terms 
due to the orthogonality of different Fock states. 

As a result, the terms proportional to $\Delta {\cal K}_i$ and ${\cal T}_i$ 
in eq.(\ref{gtd}) should be much smaller than the contribution 
from $\Delta q^T_i$, and can be reasonably neglected. Therefore, 
$g_T(x,Q^2)$ can be simply reduced to
\begin{equation}  \label{fgt}
	g_T(x,Q^2) = \sum_i e_i^2 {m^c_i+\chi_i \over 2xM} 
		~\Delta q^T_{i}(x,Q^2) ,
\end{equation}
where up to the leading log$Q^2$ QCD corrections,
\begin{equation}
	\Delta q^T_{i}(x,Q^2) = \Delta q^T_{i}(x,\mu^2) + {\alpha_s\over 
		2\pi}C_f \ln{Q^2\over \mu^2} \int_x^1 {dy\over y} P^T_{qq}
		({x\over y}) \Delta q^T_{i}(y,\mu^2)  \label{eve}
\end{equation}
with
\begin{equation}  \label{evgt}
	P^T_{qq}(x) = {1+2x-x^2 \over (1-x)_+} +{1\over 2}\delta(1-x) ,
\end{equation}
which is obtained from eq.(\ref{ems}-\ref{gtg}). 

The physical picture of $g_T$ is clear now. It probes 
the helicity flip interactions in hadrons on the 
light-front. Its dominant part is proportional to the transverse 
polarized parton distribution so that it has the well-defined
parton picture.  Since parton distributions are manifestation of
the nonperturbative QCD dynamics and helicity 
flip on the light-front describes chiral symmetry breaking,  
the structure function $g_T$ indeed directly measures 
the QCD dynamics of chiral symmetry breaking. 
We can determine this chiral 
symmetry breaking effect in $g_T$ by introducing the parameter 
$\chi_i$ which is of the order $\Lambda_{QCD}$.  This physical 
picture is extracted 
from the dominant contributions of the quark-gluon interactions
by analyzing the hadronic state in terms of Fock space wavefunctions 
on the light front. Such an analysis is extremely difficult to 
perform in the standard operator product expansion method.  

To examine this picture, we shall next compute $g_2$. 
By directly calculating $g_1(x,Q^2)$ up to the leading
log$Q^2$, we have
\begin{equation}  \label{fg1}
	g_1(x,Q^2) = \sum_i {e_i^2 \over 2}~\Delta q^L_{i}(x,Q^2),
\end{equation}
where  $\Delta q^L_{i}(x,Q^2)$ satisfies the same form of eq.(\ref{eve})
but $P_{qq}^T(x)$ is replaced by 
\begin{equation}  \label{evg1}
	P_{qq}(x) = {1+x^2 \over (1-x)_+} +{3\over 2} \delta(1-x).
\end{equation}
In both eqs.(\ref{fgt}) and (\ref{fg1}), we have not included the 
possible contributions from polarized gluon distributions.

We can now extract $g_2$ from our results of $g_1$ and $g_T$, 
\begin{equation}
	g_2(x,Q^2) = \sum_i {e_i^2\over 2} \Bigg\{ {m^c_i+\chi_i
	\over xM} \Delta q^T_{i}(x,Q^2) - \Delta q^L_{i}(x,Q^2)\Bigg\}.
\end{equation}
Although $\Delta q^T_{i}(x,Q^2)$ may not be the same as $\Delta 
g^L_{i}(x,Q^2)$ since their scale evolution functions are 
different [see eqs.(\ref{evgt}) and (\ref{evg1})], if we would
approximately take $\Delta q^T_{i}(x,Q^2) \simeq \Delta q^L_{i}(x,Q^2)$,
then we have
\begin{equation}
	xg_2(x,Q^2) \simeq \Bigg({\chi \over M} - x\Bigg) 
		g_1(x,Q^2).  \label{sg2}
\end{equation}
Here, we have ignored the current quark mass and taken $\chi$
the average value of the $u$ and $d$ quarks. Eq.(\ref{sg2})
is just our {\it oversimplified estimate} for $g_2$. We should 
also emphasize that the above result has nothing to do with the 
Wandzura-Wilzeck relation.  Taking approximately 
$\chi \simeq 200$ MeV, then ${\chi \over M}\simeq {1\over 5}$. 
Since $g_1$ has been accurately measured\cite{e143g1}, we can
estimate $g_2$ from the above equation, and find that the 
result agrees very well with the current experimental data of 
$g_2$, as shown in Fig.1. 

In conclusion, we have explored the transverse polarized structure 
function in DIS in terms of QCD and the hadronic bound state 
structure on the light-front. We find that the dominant 
contributions to transverse polarized structure function $g_T$
from all the sources, the quark mass, the transverse quark 
momentum and the quark-gluon coupling, originate from 
the chiral symmetry breaking interactions in light-front 
QCD, and they are proportional to transverse polarized 
parton distribution. The interference effects from the transverse 
quark momentum and the quark-gluon coupling in eq.(\ref{gtd}) 
are less important at high $Q^2$. As a result of the 
nonperturbative QCD dynamics of chiral 
symmetry breaking, we would expect that the magnitude of $g_T$  
is close to that of $g_1$ at high $Q^2$, namely, $g_2=g_T-g_1$ 
is very small. If the chiral symmetry breaking would 
not play the dominant role in $g_T$, one would have a small
value for $g_T$ so that $g_2$ would be close to $-g_1$.  Thus, 
further experimental measurements of $g_T$ at high $Q^2$ can 
provide a precise test of the relation between $g_T$ and the 
dynamical chiral symmetry breaking proposed in this work.   

\acknowledgments
We acknowledge useful discussions with S. J. Brodsky, X. Ji, 
J. W. Qiu, J. Ralston, L. Susskind and J. Vary. 
This work is supported in part by the U. S. Department of 
Energy under Grant No. DEFG02-87ER40371.

\begin{figure}
\caption[ ]{The prediction is extracted from the $g_1^p$ data  
\cite{e143g1} using eq.(\ref{sg2}). The $g_2^p$ data is from 
SLAC E143 \cite{E143}.}
\end{figure}

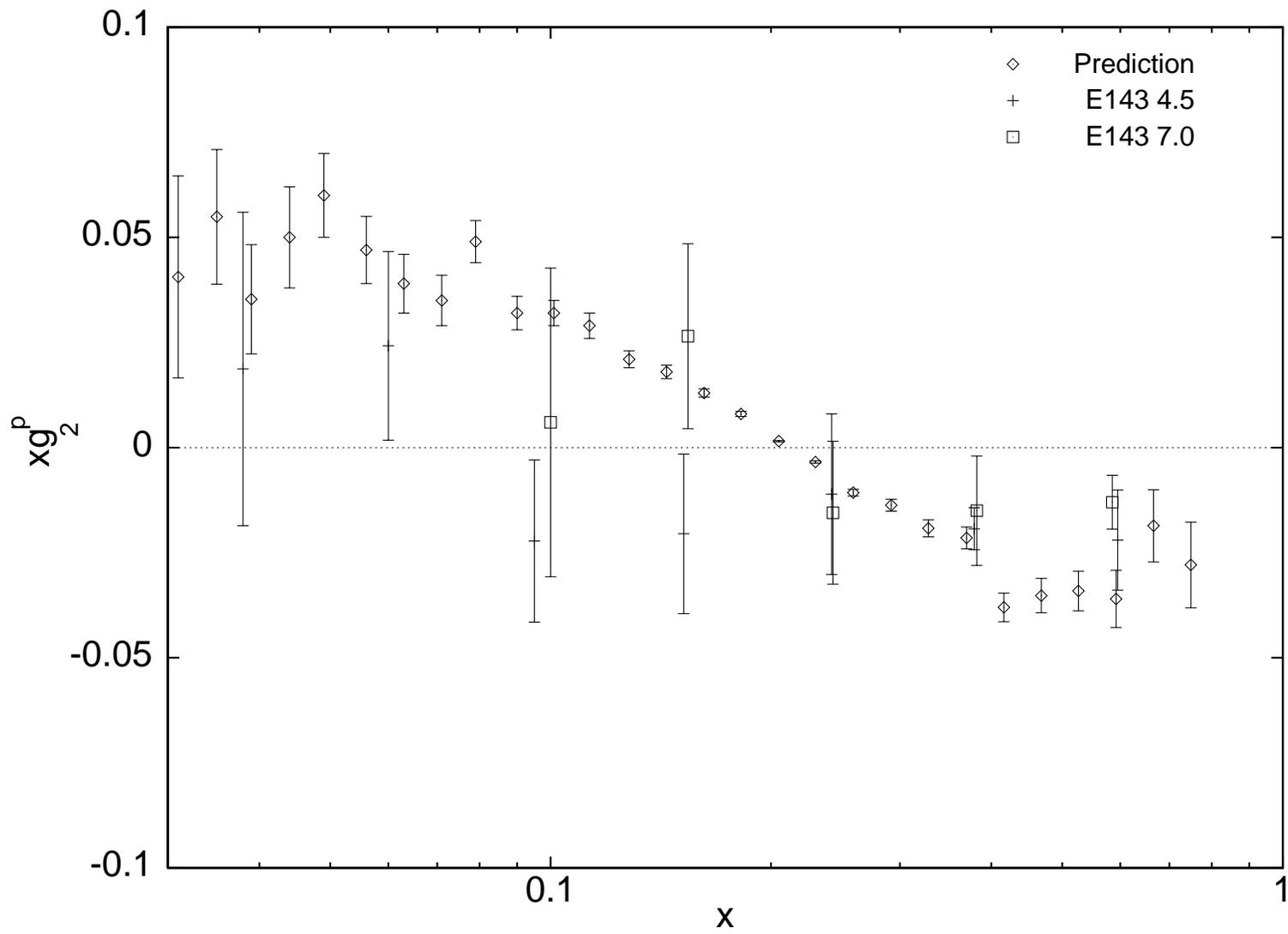
\end{document}